\documentclass[12pt]{iopart}

\usepackage{multicol}

\usepackage{graphicx}

\begin{document}

\title[Unitary model for  atomic ionization  by  intense XUV laser pulses]{Unitary model for  atomic ionization  by an intense XUV laser pulses}

\author{M.G. Bustamante and V.D. Rodr\'iguez}

\begin{abstract}
A unitary model describing the electronic transitions in  an atom subject to a strong high frequency laser pulse is proposed. The model fully accounts for the initial state coupling with the continuum spectrum.  Continuum-continuum  as well as discrete-discrete transitions are neglected. The model leads to a single integro-differential equation for the initial state amplitude.  Exact numerical and approximate closed semi-analytical solutions of this equation are obtained. A comparison of present results with full time dependent Schr{\"o}dinger equation solution for Hydrogen atoms subject to a laser pulse is presented. The initial state time dependent population is rather well described by the model and two  approximate solutions. The electron energy spectrum is also well
 reproduced by the model and by a new improved Weiskopf-Wigner related approximation.
\end{abstract}

\address{Departamento de F\'{\i}sica and Instituto de F\'{\i}sica de Buenos Aires, FCEyN,
 Universidad de Buenos Aires, 1428 Buenos
Aires Argentina}


\section{Introduction}
In recent years a variational method for describing laser-atom interactions has been proposed. The so called modified Coulomb-Volkov (MCV2$^{-}$) theoretical approximation is based on this two-front approach to provide the ionization amplitudes for atomic multiphoton ionization~\cite{Rodri04}. For the final state, the Coulomb-Volkov wave function is used. This wave function accounts for continuum-continuum coupling. For the initial state, previous options were either the simple unperturbed wave function or some	 expansions in terms of intermediate transient states~\cite{Rodri04}.
However, for non-perturbative situations, depletion of the initial state  should be considered.
In this work we introduce a model accounting for full initial state coupling with the continuum. Any other  discrete-discrete  or discrete-continuum or continuum-continuum transitions are neglected.\\
The discrete state coupled to a continuum has been mostly studied in the context of a time independent perturbation theory \cite{Fano61}.
A clear exposition of this treatment has been done by Cohen-Tannudjii \cite{Cohen92}. Here we would like to work with the same \emph{Unitary model for atomic ionization by intense XUV laser pulses} two basic ideas but without any particular assumptions with respect to the initial state-continuum coupling structure. We focus here on the time-dependent problem posed by a short laser pulse interacting with matter. Although a new kind of Weiskopf-Wigner approximation is also derived, we would like to asses the model by numerically solving the single integro-differential equation obtained below. A comparison with the full numerically solved time-dependent Schr\"odinger equation (TDSE)is performed to obtain the  survival probabilities as well as  the electron ionization spectrum.\\
There has been recent work on these subjects \cite{Gayet05,Gayet06}. However, these authors have worked only  within the  Weiskopf-Wigner approximation.  They formulate a renormalized Coulomb-Volkov approximation RCV2 \cite{Gayet05}, and also a reduced set of coupled state equations RSCSE \cite{Gayet06}. Our approach is closely related to this  work as we formulate the model as standalone, and study its performance to explain both the initial state population as well as the electron spectrum, without using either a Coulomb-Volkov wave function or the variational principle. However, in the present work we have actually solved the single integro-differential equation for the initial-state amplitude. Further, a new improved Weiskopf-Wigner semi-analytical approximation for solving this equation is obtained. Comparison of time-dependent survival probabilities for the model with full TDSE results has also been made.\\
Some authors  \cite{Paspalakis98} have attempted to improve on the Weiskopf-Wigner approximation by taking into  account counter rotating terms in the rotating wave approximation. Here we are interested in short laser pulses with only few cycles and therefore no rotating wave approximation can be used. The main component of the model is a direct population of the continuum from the initial state. Therefore we consider a laser pulse frequency larger than the ionization potential. On the other hand, full treatment of the initial-state coupling with the continuum  is allowed. The simplicity of the model is appealing because once the time-dependent initial state amplitude has been determined, transition amplitudes to all the other states included in the model are straightforwardly established. The sum of the transition probabilities to all the states included in the model remains unity.  We have worked in the length gauge.\\
In section II, the  formulation of the model is presented.  In section III,
the results and their comparison   with a TDSE calculation for hydrogen under a XUV laser pulse are analyzed. In Sec. IV, some conclusions and future insights to be taken into account are put forward. The proof of the important unitary property of this model is given in the Appendix. Atomic units are used unless otherwise stated.

\section{Theory}

\subsection{The model}

In the one-active electron approximation, the laser-atom Hamiltonian  in the length gauge, is given by
\begin{equation}
\qquad \qquad \qquad H(t)=H_{a} + \mathbf{r} \cdot \mathbf{F}(t)
\label{AtomicHamiltonian1}
\end{equation}
where
\begin{equation}
\qquad \qquad \qquad H_{a}=-\frac{\nabla^{2}}{2}+V_{a}(r)
\label{AtomicHamiltonian2}
\end{equation}
is the non relativistic atomic Hamiltonian in the absence of the laser with atomic potential  $V_{a}(r)$. In the laser-atom interaction term, $\mathbf{r}$ is the position of the electron and $\mathbf{F}(t)$ the laser pulse electric field.
  The time dependent Schr\"odinger equation (TDSE) in the dipolar approximation is then given by
\begin{equation}
\qquad \qquad i\frac{\partial{| \psi (t)\rangle }}{\partial{t}} =[H_a + \mathbf{r} \cdot \mathbf{F}(t)]| \psi (t)\rangle .
\label{SEq}
\end{equation}
We perform the expansion of the time dependent state ket of the system in  eigenkets   of the atomic Hamiltonian, separating the initial state  $|i\rangle$ with energy $\varepsilon_{i}$, thus
\begin{equation}
\fl
\qquad\qquad\qquad|\psi(t)\rangle = a_{i}(t)\exp[-i\varepsilon_{i}t] |i\rangle +  \int d\mathbf{k} a_{\mathbf{k}}(t)\exp[-i\varepsilon_{k}t]|\mathbf{k}\rangle,
\label{StateKetExpansion}
\end{equation}
where $|\mathbf{k}\rangle$ is the eigenket corresponding to the continuous $\varepsilon_{k}$ energy level, respectively. The momentum eigenkets are assumed to be normalized in the momentum scale.	
When the last expansion is replaced into the TDSE and projected onto each ket considered in the expansion, the following  system of first order coupled  differential equations is obtained:
\begin{eqnarray}
\fl
\qquad\qquad\qquad\qquad \dot{a}_{i}(t)=-i \int d\mathbf{k}V_{i\mathbf{k}}(t) \exp[-i(\varepsilon_{k}-\varepsilon_{i})t]a_{\mathbf{k}}(t)\label{DiAmplitude}
\end{eqnarray}
\begin{eqnarray}
\fl
\qquad\dot{a}_{\mathbf{k}}(t)=-i V_{\mathbf{k}i}(t) \exp[-i(\varepsilon_{i}-\varepsilon_{k})t]a_{i}(t)
-i \int d\mathbf{k^{\prime}}V_{\mathbf{k}\mathbf{k^{\prime}}}(t)\exp[-i(\varepsilon_{k^{\prime}}-\varepsilon_{k})t]a_{\mathbf k^{\prime}}(t)\label{Dk0Amplitude}
\end{eqnarray}
Our main approximation consists in keeping only the terms coupling  the initial state with the continuum part of the atomic spectrum. Therefore, the
 modified (\ref{Dk0Amplitude}) reads
\begin{eqnarray}
\fl
\qquad\qquad\qquad\qquad \dot{a}_{\mathbf{k}}(t)=-i V_{\mathbf{k}i}(t) \exp[-i(\varepsilon_{i}-\varepsilon_{k})t]a_{i}(t)\label{DkAmplitude}
\end{eqnarray}
\begin{figure}
\includegraphics[width=12 cm,height=8 cm]{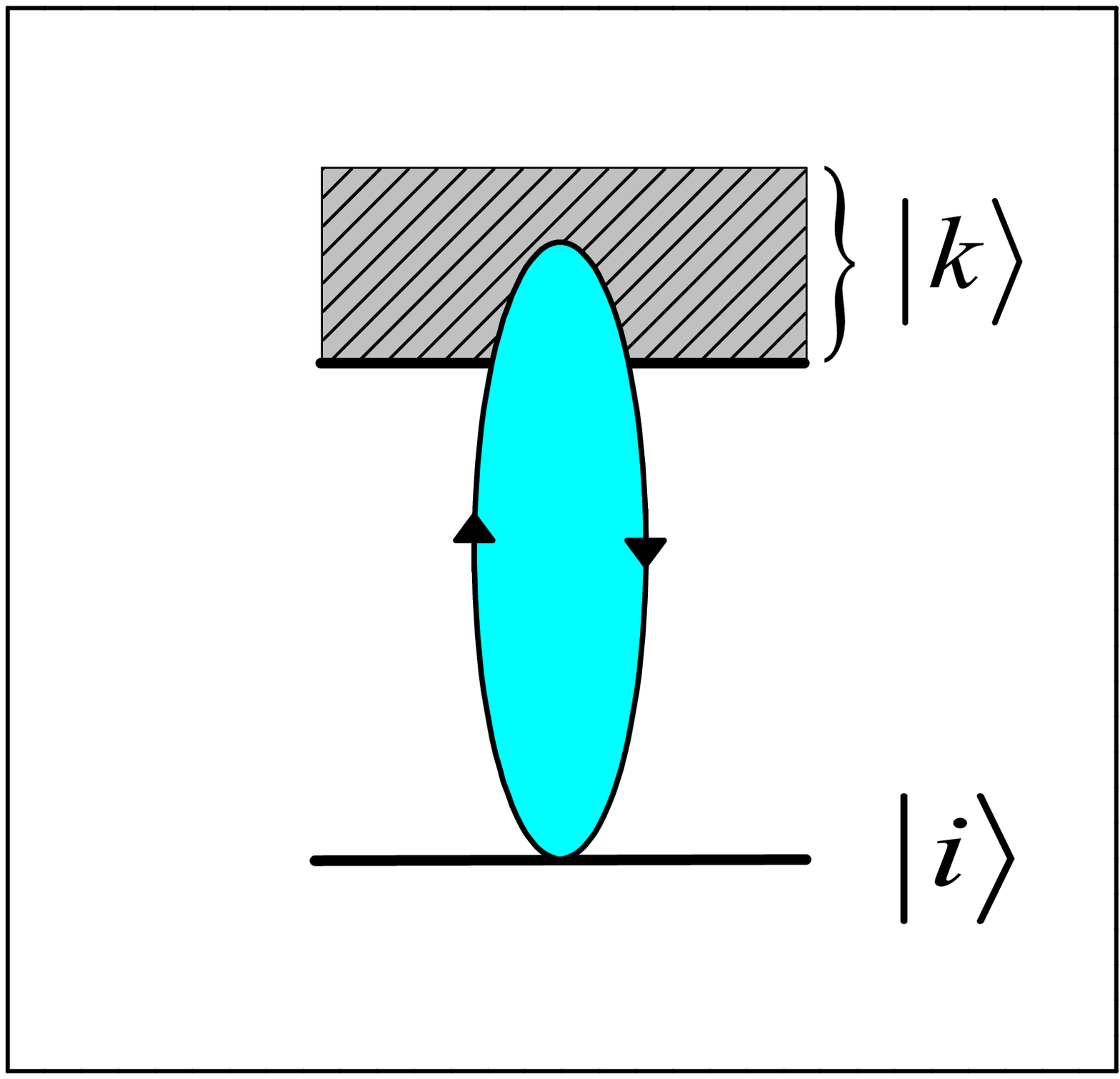}
\begin{caption}{Diagram for the approximations considered where only transitions between the initial state and the continuum states (used in the Weiskopf-Wigner approximation in [2]) are taken into account}
\end{caption}
\end{figure}
Formal integration of (\ref{DkAmplitude}) provides the transition amplitude to the $\mathbf{k}$-continuum states, in terms of the initial state amplitude:
\begin{equation}
\qquad a_{\mathbf{k}}(t)=-i\int_0^{t}dt^{\prime} V_{\mathbf{k}i} \exp[-i(\varepsilon_{i}-\varepsilon_{k})t^{\prime}]a_{i}(t^{\prime}),
\label{kAmplitude}
\end{equation}
Without loss of generality we take the field as linearly polarized along the $z$-direction.
Thus, 	for this laser field	polarization, $V_{\mathbf{k}i}= z_{\mathbf{k}i}=\left\langle \mathbf{k}|z|i\right\rangle$.
By replacing (\ref{kAmplitude}) in (\ref{DiAmplitude}), the following integro-differential	equation (IDE) for the initial state  amplitude is obtained:
\begin{eqnarray}
\dot{a}_{i}(t)=(-i)^{2}\int_{0}^{t}dt^{\prime}F(t)F(t^{\prime})
\int d\mathbf{k}|z_{\mathbf{k}i}|^{2}\exp[-i(\varepsilon_{k}-\varepsilon_{i})(t-t^{\prime})]
a_{i}(t^{\prime}).
\label{ai-IDE0}
\end{eqnarray}
This equation may be written as
\begin{equation}
\dot{a}_{i}(t)=(-i)^{2}\int_{0}^{t}dt^{\prime }F(t)F(t^{\prime})\exp[i\varepsilon_{i}(t-t^{\prime})]h(t-t^{\prime})a_{i}(t^{\prime}),
\label{ai-IDE1}
\end{equation}
with the kernel function
\begin{equation}
\qquad\qquad h(t-t^\prime)=\int d\mathbf{k}|z_{\mathbf{k}i}|^{2}\exp[-i\varepsilon_{k}(t-t^{\prime})].
\label{h}
\end{equation}
The diagram in figure 1 shows the main approximation considered in this work. The coupling scheme accounts for  transitions between  initial  and  the continuum states. Therefore the function $h(t-t^{\prime})$ accounts for the coupling between the initial state and the continuum  part of the atomic spectrum.
For atomic Hydrogen in the initial ground state, the function $h(t-t^{\prime})$ can be reduced to the integral
\begin{equation}
\qquad\qquad h(t-t^{\prime})=\ \int dk\exp [-i\varepsilon_{k}(t-t^{\prime})]\beta_{0}(k),
\label{h1}
\end{equation}
where
\begin{equation}
\beta_{0}(k)=\int d\Omega_k k^{2}|z_{\mathbf{k}i}|^{2}=\frac{128\exp[\frac{\pi-4\arctan(k)}{k}]k\sinh^{-1}(\frac{\pi}{Z})}{3(1+k^{2})^{5}}.
\label{beta0}
\end{equation}
In the last equation we have used the dipolar  matrix element obtained with the Nordsieck method \cite{nordsieck54} :
\begin{equation}
\qquad z_{\mathbf{k}i}= \frac{4 \sqrt {2} \exp[\frac{\pi}{2k}-\frac{2\arctan k}{k}](i+\frac{1}{k})k\cos[\vartheta_k]\Gamma[1-\frac{i}{k}]}{(1+k^{2})^{3}\pi}
\label{zki}.
\end{equation}
The integral in (\ref{h1}) has been evaluated using methods of  integration in the complex plane. The resulting function $h(t-t^{\prime})$ is a smooth function peaked near $t-t^{\prime}=0$.
The equation for the amplitude of the initial state, given by  (\ref{ai-IDE1}) together with the initial condition $a_{i}(0)=1$ define an IDE of the Volterra type. Its solution  has been obtained numerically using an algorithm based on the Taylor expansion, proposed in \cite{goldfine77}.
A Filon's like algorithm for dealing with highly oscillating functions integrals \cite{milovanovic77} has been adapted
for computation of the different terms of the IDE.\\
Because of the selection rules for dipolar transition, only states with orbital quantum number $l=1$, coupled with the $1s$ initial state contribute to the expansion of the system ket state.
Once the numerical solution of the single IDE has been obtained, transition amplitudes to continuum states are readily worked out using (\ref{kAmplitude}).

\subsection{Ionization rate}
The total ionization probability is obtained by integrating the square modulus of transition amplitude to the continuum	 with momentum $\mathbf{k}$,  over the momentum space,
\begin{equation}
\qquad\qquad\qquad P_{ioni}(t)=\int d\mathbf{k}|a_{\mathbf{k}}(t)|^{2}.
\label{Pioni}
\end{equation}
The  \emph{ionization rate} is then given by
\begin{equation}
\qquad\qquad \dot{P}_{ioni}(t)=\int d\mathbf{k} a_{\mathbf{k}}(t)\dot{a}_{\mathbf{k}}^*(t)   +  \texttt{c.c.}
\label{DPioni}
\end{equation}
where $\texttt{c.c.}$ denotes the complex conjugate of the first term in the right hand side.
Replacing the expression for  $a_{\mathbf{k}}(t)$ given by (\ref{kAmplitude}), after performing the corresponding temporal derivative, we obtain:
\begin{eqnarray}
\fl
\qquad \dot{P}_{ioni}(t)=\int_{0}^{t}dt^{\prime}F(t)F(t^{\prime})\exp[i\varepsilon_{i}(t-t^{\prime})]a_{i}^*(t)a_{i}(t^{\prime}) \int d\mathbf{k}|z_{\mathbf{k}i}|^{2}\exp[-i\varepsilon_{k}(t-t^{\prime})] \nonumber \\
\qquad\qquad\qquad\qquad\qquad + \texttt{c.c.}
\end{eqnarray}
Using (\ref{h}) the ionization rate may be expressed as
\begin{equation}
\fl
\qquad\qquad \dot{P}_{ioni}(t)=F(t)\exp(i\varepsilon_{i}t)a_{i}^*(t)\int_{0}^{t}dt^{\prime }F(t^{\prime})\exp[-i\varepsilon_{i}t^{\prime}]a_{i}(t^{\prime})h(t-t^{\prime})+\texttt{c.c.}
\label{DPioni1}
\end{equation}
The proof of the unitary property of the model based on the ionization rate given in (\ref {DPioni1}) is given in appendix A.
The time-dependent total ionization probability is obtained by a simple numerical integration of this differential equation,  bearing in mind the initial condition $P_{ioni}(0)=0$.

\subsection{Approximated analytical solutions}

The kernel $h(t-t^{\prime})$ defined in (\ref{h}) is peaked near $t^{\prime}\sim t$. By considering a smooth temporal behavior for the initial state amplitude  we may use the Taylor  expansion of the amplitude $a_{i}(t^{\prime})$ to order zero by taking	 $a_{i}(t^{\prime})\sim a_{i}(t)$. In that case the IDE given in (\ref{ai-IDE1}) becomes the first order ODE
\begin{equation}
\qquad\qquad\qquad\qquad \dot{a}_{i}(t)=-i\alpha(t)a_{i}(t),
\end{equation}
whose exact solution is,
\begin{equation}
\qquad\qquad\qquad a_{i}(t)=\exp[- i\int_{0}^t dt^{\prime\prime}\alpha(t^{\prime \prime})],
\label{ODE10Sol}
\end{equation}
with the function $\alpha(t)$ defined as
\begin{equation}
\fl
\qquad\qquad\qquad\qquad \alpha(t)=-i\int_{0}^{t}dt^{\prime }F(t)F(t^{\prime})\exp[i\varepsilon_{i}(t-t^{\prime})]h(t-t^{\prime}).\\
\label{AlphaFunc}
\end{equation}
This  approximated amplitude of the initial state is related to the well known Weiskopf-Wigner approximation and has been recently used in \cite{Gayet05} to compute the Hydrogen ionization spectra by laser pulses.\\
Now we improve this approximation  by using the first order Taylor expansion $a_{i}(t^{\prime})\sim a_{i}(t)+\dot a_{i}(t) (t^{\prime}-t) $, obtaining the first order ODE
\begin{equation}
\qquad\qquad\qquad (1+\beta(t))\dot{a}_{i}(t)=-i\alpha(t)a_{i}(t),
\end{equation}
The solution of this differential equation is
\begin{equation}
\fl
\qquad\qquad\qquad\qquad\qquad\qquad a_{i}(t)=\exp\left[-i\int_{0}^{t}dt^{\prime\prime}\frac{\alpha(t^{\prime\prime})}{1+\beta(t^{\prime\prime})}\right],
\label{ODE1Sol}
\end{equation}
where the function $\beta(t)$ is defined by
\begin{equation}
\fl
\qquad\qquad\qquad\qquad \beta(t)=-\int_{0}^{t}dt^{\prime }F(t)F(t^{\prime})(t^{\prime}-t)\exp[i\varepsilon_{i}(t-t^{\prime})]h(t-t^{\prime}).
\label{BetaFunc}
\end{equation}

\section{Results}

We present the results obtained for atomic Hydrogen under an intense XUV laser pulse. The laser frequency is taken to be larger than the ionization potential.
The electric field pulse is defined by
\begin{equation}
\qquad \mathbf{F}(t)=
\left\{
\begin{array}{cc}
\mathbf{F}_{0}\sin(\omega t+\varphi) \sin^{2}(\frac {\pi t}{\tau}) \qquad & \texttt{if} \qquad t\in{(0,\tau)}\\
0 \qquad & \texttt{elsewhere},
\end{array}
\right.
\label{Pulse}
\end{equation}
and it is linearly polarized along the $z$-direction: $\mathbf{F}_{0}=F_{0}\widehat{z}$. The parameter $\tau$ stands for the duration of the pulse, the sine-square factor defines the pulse envelope, $\omega$ is the laser angular frequency and $\varphi$ the phase of the carrier. We fix this phase to have the maximum of the pulse at half cycle.\\
In what follows the laser field frequency is taken to be $\omega=0.6$ a.u. and a	$20$ cycles pulse length is considered.
We focus here on  the module square of the initial state amplitude.
In figure 2, two field amplitudes ($0.05$ a.u. and $0.1$ a.u.) are considered.
The scattered plots give the results provided by the TDSE Qprop code \cite{bauer06}.
This calculations are shown only for times $t_i$ such that $A(t_i)=0$. Note that comparison between results obtained in the length gauge (UM)with those  obtained in the velocity gauge (QPROP) is meaningless for  times other than $t_i$ \cite{ChiDong}. The survival probability corresponding to the UM approximation is in very good agreement with the TDSE solution  for all  the pulse duration.\\
\begin{figure}
\includegraphics[width=14 cm,height=9 cm]{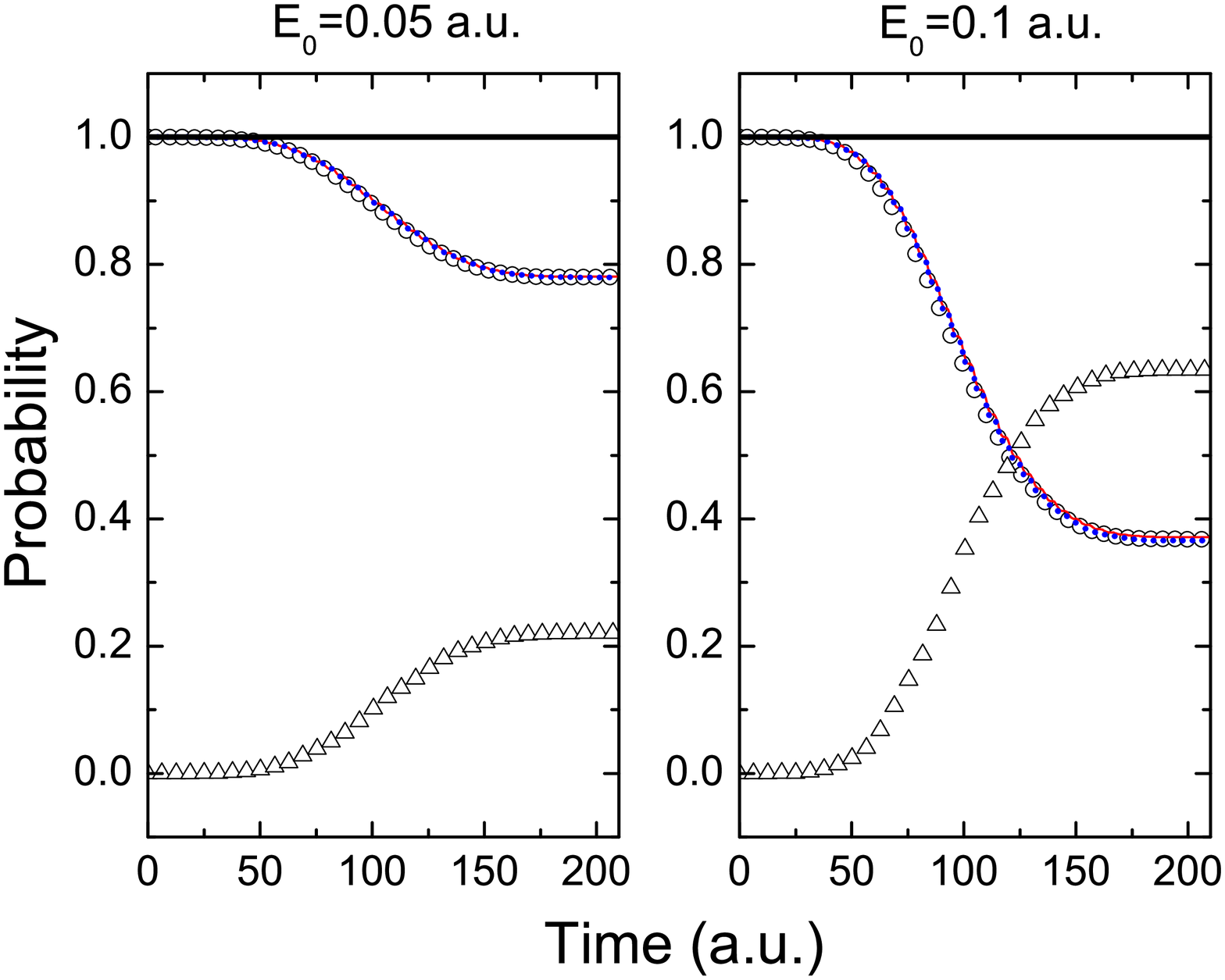}
\begin{caption}{Survival, total ionization   and  sum of all transition probabilities for a laser pulse of angular frequency of $\omega=0.6$ a.u. and $20$ cycles. Two cases are considered for the laser electric field amplitudes: $E_0=0.05$ a.u. ($8.792 \times 10^{13} W/cm^{2}$) (left) and $E_0=0.1$ a.u. ($3.517 \times 10^{14} W/cm^{2}$) (right). Circles, TDSE survival probability obtained with the Qprop code [2].
Dotted line (blue online), solution of (\ref{ai-IDE1}) . Solid line (red online), Weiskopf-Wigner approximation. Triangles, UM total ionization probability. Wide solid  line, total sum of all UM transition probabilities.}
\end{caption}
\label{ProbPhases1}
\end{figure}
The depletion of the initial state becomes significant in the second half of the pulse.
Both versions, the full model  and
its  Weiskopf-Wigner approximation (equation (21)), remain near each other along all the pulse duration and  agree with the exact survival probability.\\
In figure 3 we show the same magnitude for stronger laser fields.
We display the survival probabilities on the left side for $E_0=0.2$ a.u. and on the right side for $E_0=0.3$ a.u.
\begin{figure}
\includegraphics[width=14 cm,height=9 cm]{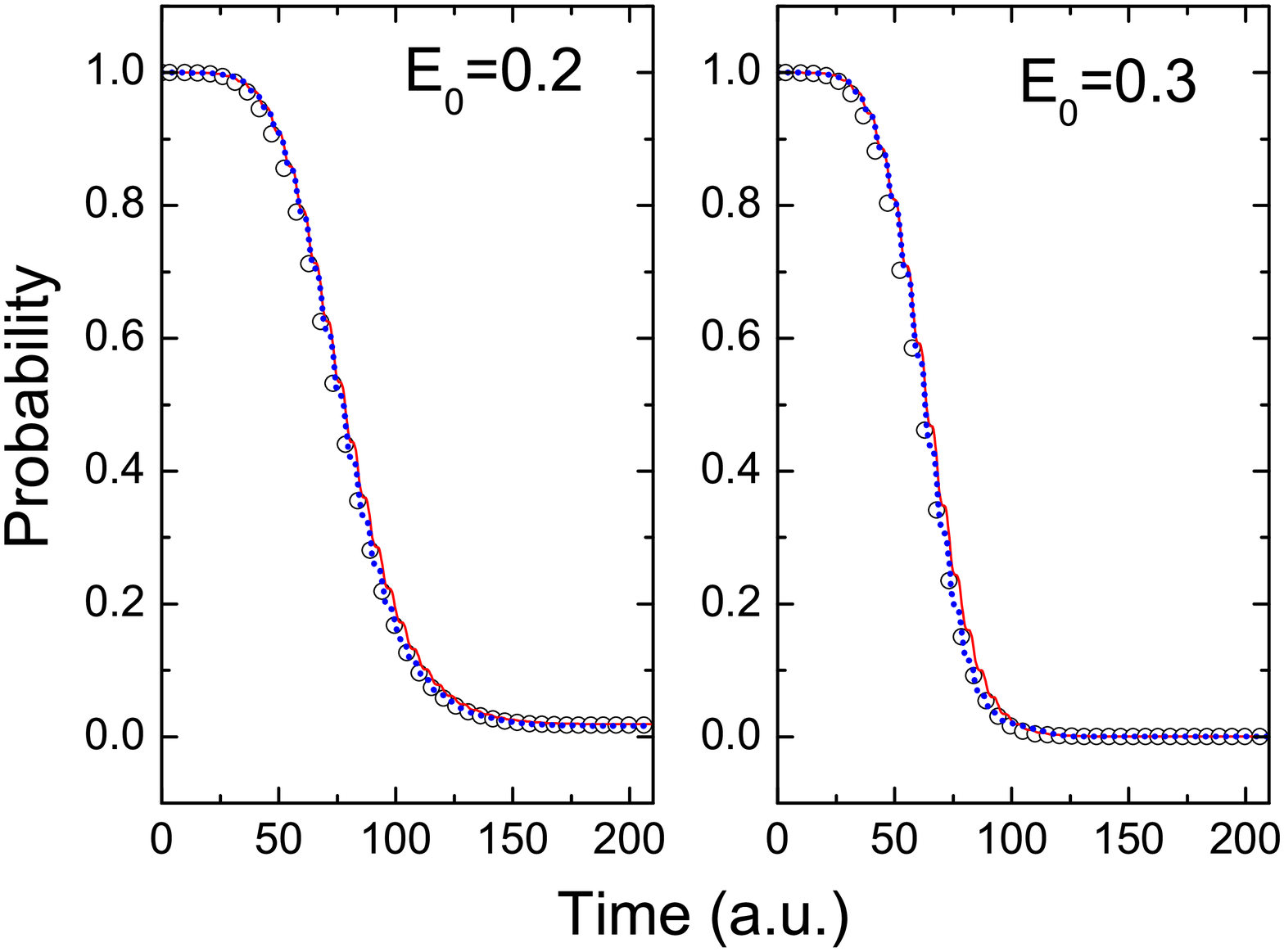}
\begin{caption}{Same as in figure 3 for higher field amplitudes: $E_0=0.2$ a.u. ($1.407 \times 10^{15} W/cm^{2}$) (left) and $E_0=0.3$ a.u. ($3.165 \times 10^{15} W/cm^{2}$) (right). In this case the total ionization and the sum of all transition probabilities are not displayed. Frequency	and number of cycles as in figure 2.}
\end{caption}
\label{ProbPhases2}
\end{figure}
The model runs closer to the exact TDSE solution even  when the initial state depletion becomes important.
The unitary property of the model is behind this success as  ionization probability does saturate in a unity value for the second half pulse and therefore, survival probability goes to zero. In this case the total ionization and the sum of all transition probabilities have not been displayed since they follow the same behavior as in figure 2.\\
Figure 4 shows the	spectra	 for the lower field intensities  $E_0=0.05$ a.u. and $E_0=0.1$ a.u..
The plots are presented in logarithmic scale in order to appreciate the	 global spectra.
The TDSE spectra account for up to four ATI peaks.
\begin{figure}
\includegraphics[width=14 cm,height=9 cm]{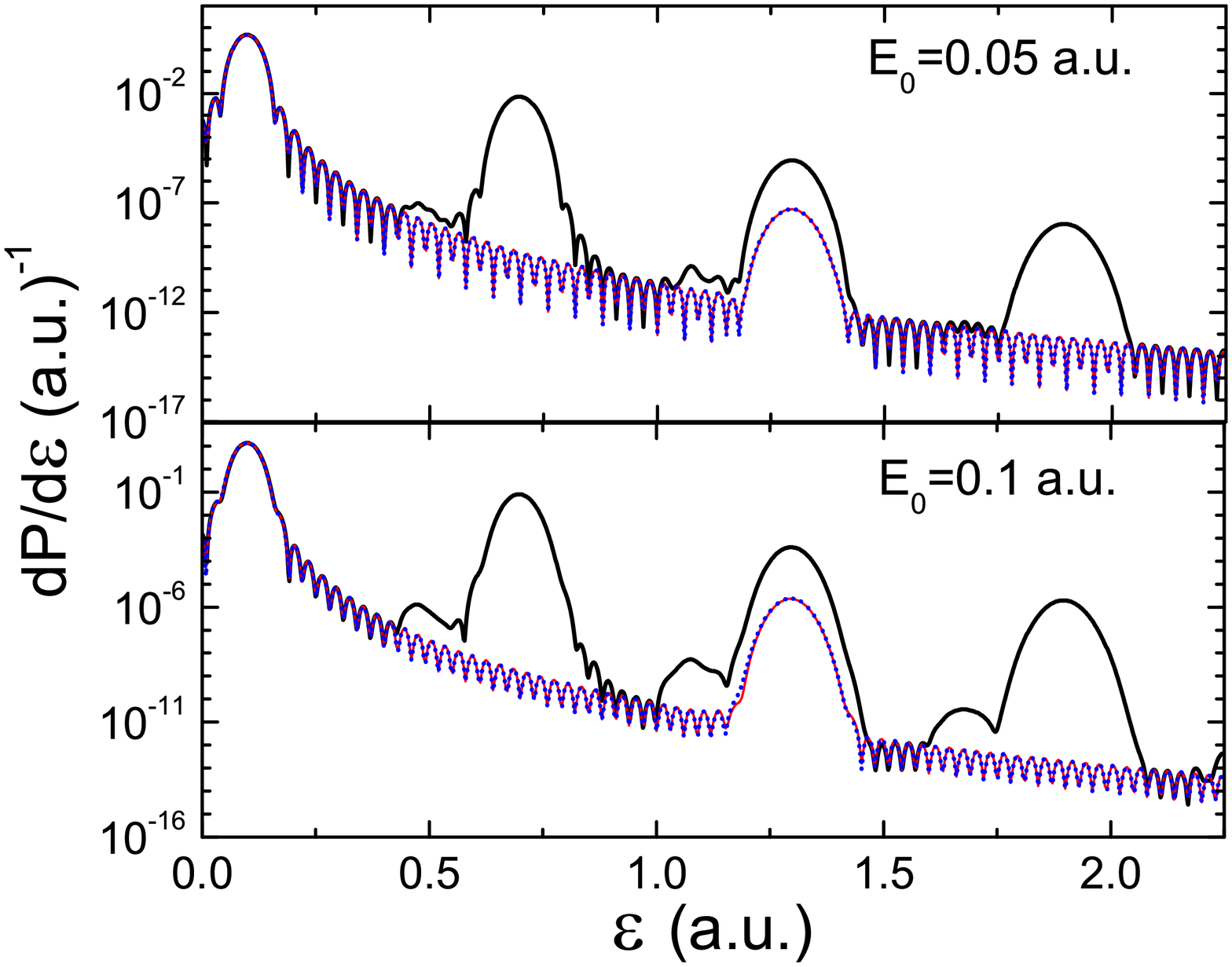}
\begin{caption}{Electron spectrum: wide solid line, exact TDSE obtained with the Qprop code \cite{bauer06};
dotted line (blue online), UM results ; thin solid line (red online), Weiskopf-Wigner . Frequency	 and number of cycles as in figure 2.}
\end{caption}
\label{Spectra1}
\end{figure}
Again, no appreciable differences are found between the Weiskopf-Wigner calculation and the exact solution of the integral equation. The model displays the first ionization peak in agreement with the one on TDSE. No second ATI peak is shown as by selection-rules this peak is populated with $s$ and $d$ electrons and therefore not accounted for in the model.\\
Recall that only $p$ states are coupled with the initial ground state. However, a small third ATI peak can be appreciated in the model results. This peak is populated by straightforward three-photon transitions from the ground state to the	 p-states in the continuum, without any real $s$ or $d$ intermediate state. The TDSE structure before the second and the third ATI peaks is due to transient intermediate state populations.
Our UM model cannot account for this structure as it corresponds to a two photon absorption from an excited level \cite{Rodri04}. The background is well reproduced by the model.\\
Figure 5 shows the same spectra but for higher electric field amplitudes $E_0=0.2$ a.u. and $E_0=0.3$ a.u. We only display the first ionization peak in a linear scale.
First of all, we may appreciate a nice agreement between our model and the TDSE results. On the other hand, the Weiskopf-Wigner approximation is larger than the exact TDSE
and its maximum position is shifted towards the lower energies. Also the figure shows the improved Weiskopf-Wigner approximation obtained by using equations (23) and (24).
This result is one of the main outcomes in this work. The semi-analytical approach has the predictive power of the full UM preserving the simplicity of the Weskopf-Wigner approximation. We should remark that no rotating wave approximation has been performed.
\begin{figure}
\includegraphics[width=14 cm,height=9 cm]{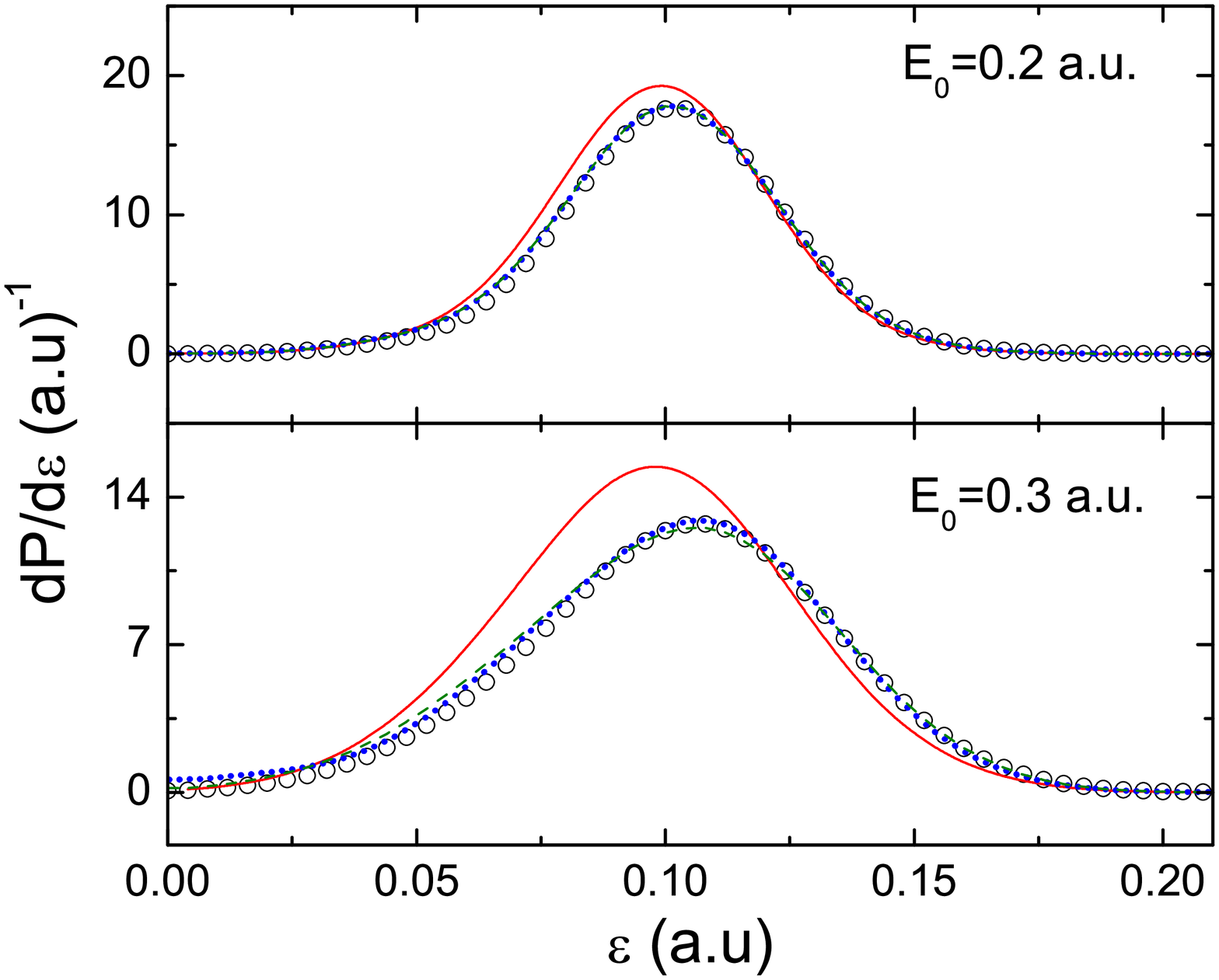}
\begin{caption}{Electron spectrum: empty circles, exact TDSE obtained with the Qprop code \cite{bauer06}; dotted line (blue online), UM results; solid line (red online), Weiskopf-Wigner approximation; dashed line (green online), improved Weiskopf-Wigner approximation obtained from (\ref{ODE1Sol}). Frequency and number of cycles as in figure 2.}
\end{caption}
\label{Spectra2}
\end{figure}
Another indication on the quality of the present approximations can be obtained by analyzing the total ionization probabilities as a function of the electric field amplitude.
Table 1 shows the total ionization probabilities. Results obtained with the Weiskopf-Wigner approximation exceed unity for $E_0>0.2$ a.u.. The UM remains closer to the QPROP results up to $E_0=0.3$ a.u.. The tiny declining trend shown by TDSE ionization probabilities between $E_0=0.3$ a.u. and $E_0=0.4$ a.u. is  also reproduced by this model. On the other hand, the improved Weiskopf-Wigner approximation model remains close to the TDSE although the unitarity is broken for the two larger field amplitudes. Although full UM is unitary, their approximations do not satisfy this property as expected.

\section{Final remarks}

A unitary model has been introduced for the theoretical study of the atom dynamics perturbed by a strong XUV laser pulse. The model leads to a single integro-differential equation that can be easily solved in a numerical way. Simple analytical expressions can be found within the Weiskopf-Wigner approximation and a new improved Weiskopf-Wigner approximation. The model is useful for high frequency laser atom ionization and for lower frequency when applied to negative ion detachment.

\ack
This work was partially supported by the  Consejo Nacional de Investigaciones Cient\'{\i}ficas y T\'{e}cnicas PIP  and Universidad de Buenos Aires UBACYT, Argentina.

\begin{table}
\caption{\label{arttype}Total ionization probabilities}
\footnotesize\rm
\begin{tabular*}{\textwidth}{@{}l*{15}{@{\extracolsep{0pt plus12pt}}l}}
\br
$E_0$(a.u.)& Weiskopf-Wigner & Improved Weiskopf-Wigner & UM & QPROP \\
\mr
0.05&	   0.220&	   0.220&	   0.220&	   0.220\\
0.10&	   0.639&	   0.635&	   0.634&	   0.632\\
0.20&	   1.047& 	   0.989&	   0.983&	   0.983\\
0.30&	   1.123&	   1.055&	   0.998&	   0.999\\
0.40&	   1.170&	   1.174&	   0.994&	   0.997\\
\br
\end{tabular*}
\end{table}

\appendix

\section{}
Here we show the unitary property of the model.
Let's define two functions $F_{i}(t)$ and $\gamma_{i}(t)$ by	
\begin{eqnarray}
F_{i}(t)=F(t)\exp(i\varepsilon_{i})\\
\gamma_{i}(t)=-\int_{0}^{t}dt^{\prime }F(t^{\prime})\exp[-i\varepsilon_{i}t^{\prime}]a_{i}(t^{\prime})h(t-t^{\prime}),
\label{AuxFunc}
\end{eqnarray}
with $h(t-t^{\prime})$ given by Eq. (\ref{h}). With these functions the ionization rate Eq. (\ref{DPioni1}) takes the form
\begin{equation}
\fl
\qquad \qquad \qquad \dot{P}_{ioni}(t)= - 2 \texttt{Re}[(F_{i}(t) a_{i}^*(t) \gamma_{i}(t))].
\label{DPioni3}
\end{equation}
In a similar way, the initial state depletion rate can be obtained using Eq. (\ref{ai-IDE1})
\begin{equation}
\fl
\qquad \qquad \qquad \dot{P}_{i}(t)=2 \texttt{Re}[F_{i}(t)a_{i}^*(t)\gamma_{i}(t)].	
\label{DPi1}
\end{equation}
By adding the last equation  with  Eq. (\ref{DPioni3}) it is clear that
\begin{equation}
\fl
\qquad \qquad \dot{P}_{i}(t) +  \dot{P}_{ioni}(t) =\dot{P}_{TOTAL}(t)= 0.
\label{DPimasSumDPnmasDPioni}
\end{equation}
This means that the total  probability to all the states $P_{TOTAL}(t)$	is constant in the whole time interval $(0,\tau)$. As $P_{TOTAL}(0)=1$, then $P_{TOTAL}(t)=1$ $\forall$ $t$ $\in$ $(0,\tau)$.
In other words, the model is \emph{unitary}.

\end{document}